\documentclass[twocolumn]{webofc}
\usepackage[varg]{txfonts}   

\usepackage[english]{babel}
\usepackage{color}
\usepackage{graphicx}
\usepackage[utf8]{inputenc}
\usepackage{caption}
\usepackage{mathtools}
\usepackage{slashed}
\usepackage{enumitem}
\usepackage{emptypage}

\begin{document}
	
\title{Status of the Fermilab Muon $g-2$ Experiment}

\author{\firstname{Paolo} \lastname{Girotti}\inst{1,2}\fnsep\thanks{\email{pgirotti@fnal.gov}}
	\firstname{on behalf of the} \lastname{Muon $g-2$ (E989) Collaboration}
}

\institute{
	University of Pisa
	\and
	INFN Pisa
}

\abstract{%
	The muon anomalous magnetic moment, $a_\mu = (g-2)/2$, is a low-energy observable which can be both measured and computed with very high precision, making it an excellent test of the Standard Model (SM) and a sensitive probe of new physics. Recent efforts improved the precision of both the theoretical prediction and the experimental measurement. On the theory side, the Muon $g-2$ Theory Initiative, an international team of more than 130 physicists, reached in 2020 a consensus on the SM prediction for $a_\mu$. On the experimental side, the E989 Muon $g-2$ Collaboration at Fermilab (FNAL) published in April 2021 a new measurement of $a_\mu$ from the Run-1 dataset (2018) with 0.46 ppm precision, corroborating the previous Brookhaven National Laboratory (BNL) measurement and increasing the discrepancy with the SM value to 4.2 standard deviations. In this paper, I will discuss the experimental setup and report on the current status of the experiment.
}

\maketitle

\section{Introduction}
\label{sec:intro}

The study of the lepton magnetic moments has a long history starting back in 1928 when Dirac predicted $g=2$ for the electron. The 1948 calculation by J. Schwinger of the first-order correction due to virtual particles marked the beginning of a long sequence of improvements in both the theoretical prediction and the experimental measurement of the electron and muon anomalies. While the electron $g-2$ measurements represent a striking confirmation of QED theory, the muon anomaly has emerged as an important test of the entire Standard Model due to significant contributions from weak and strong interactions.\\
In April 2021, the Muon $g-2$ Collaboration released their Run-1 measurement of the muon anomaly with a precision of 460 ppb \cite{PRL}. This result is in agreement with the previous experiment at Brookhaven National Laboratory (E821), which determined $a_\mu$ with an accuracy of 540 ppb \cite{BNL}. By combining the two measurements into a new world average, we find a discrepancy of 4.2 $\sigma$ with respect to the current consensus of the Standard Model prediction \cite{WP2020}.
The current theoretical and experimental values are:
\begin{align}
	a_\mu^{Th}\;\;[2020] &= 116\,591\,810(43) \times 10^{-11} \textnormal{ (0.37 ppm)}\\
	a_\mu^{Exp}\,[2021] &= 116\,592\,061(41) \times 10^{-11} \textnormal{ (0.35 ppm)}\\
	a_\mu^{Exp} - a_\mu^{Th} &= (251\pm59) \times 10^{-11} \textnormal{ (4.2$\sigma$)}
\end{align}
The history of $g-2$ is far from being completed, as the Muon $g-2$ Experiment at Fermilab will improve its precision by a factor of four, other experimental efforts are in the works, and new calculations with Lattice QCD techniques represent a new way to improve on the theoretical predictions.
\begin{figure}[!ht]
	\centering
	\includegraphics[width=0.9\linewidth]{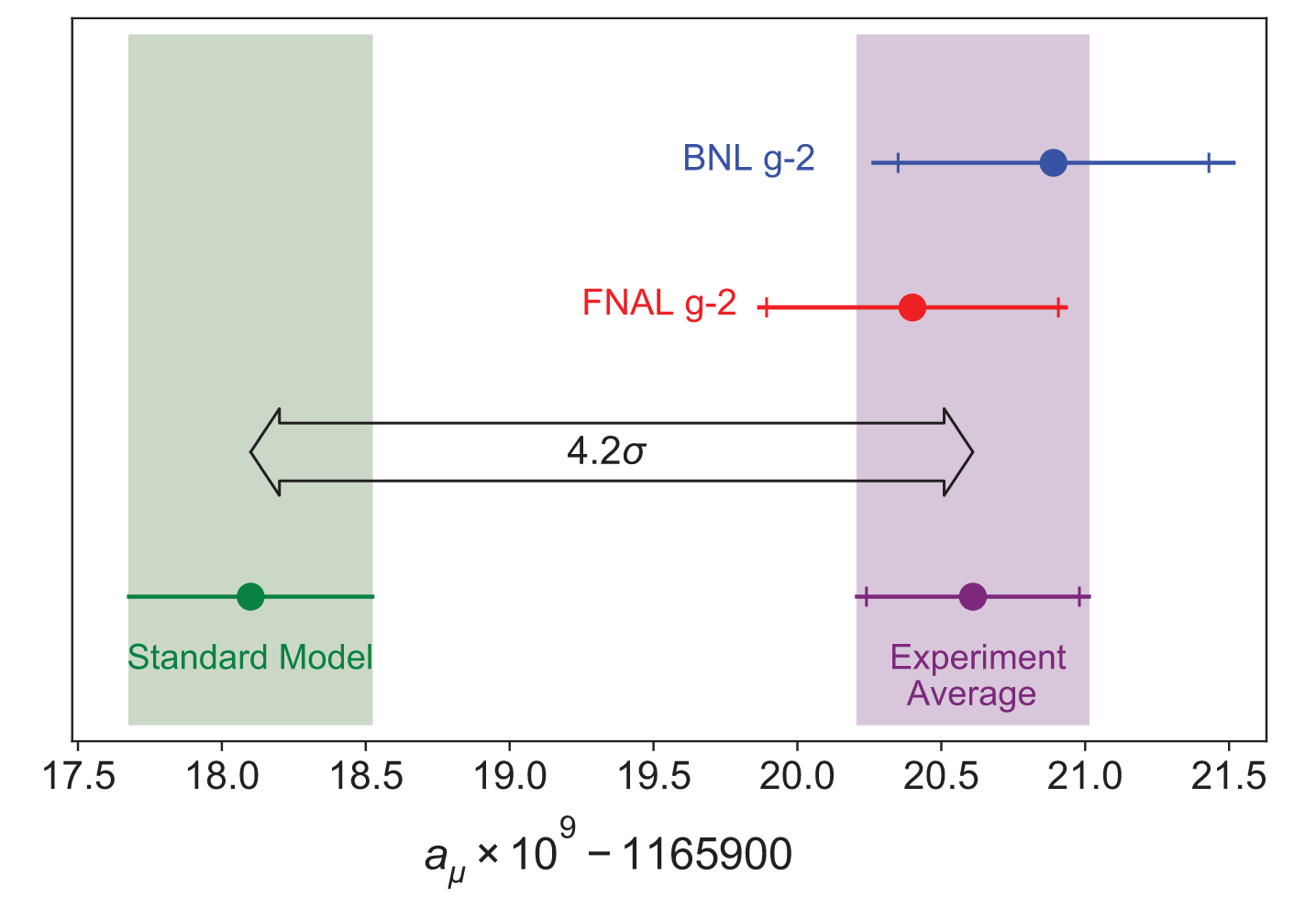}
	\caption{Run-1 FNAL measurement compared with the previous BNL experiment and the 2020 Standard Model consensus recommended by The Muon $g-2$ Theory Initiative \cite{WP2020}. Figure from \cite{PRL}.}
	\label{fig:Run-1}
\end{figure}\
\subsection{The importance of g-2}\label{sec:g-2}
\begin{figure}[!ht]
	\centering
	\includegraphics[width=\linewidth,clip]{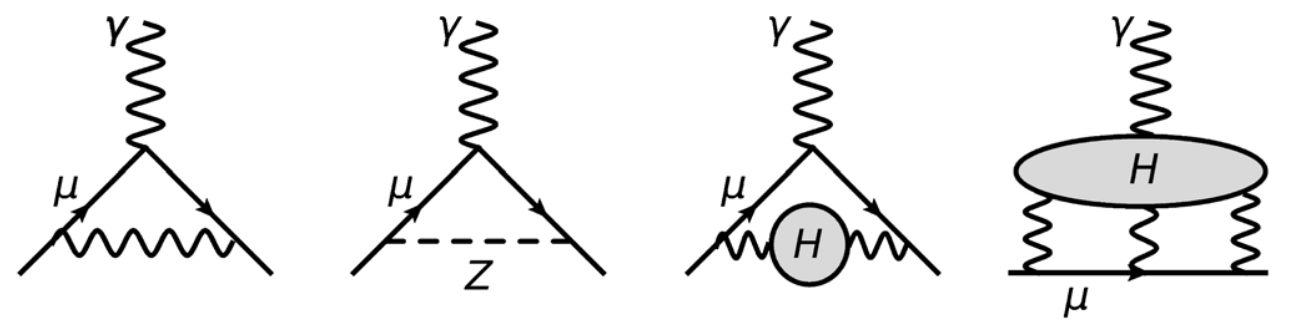}
	\caption{Lower-order examples for EM, weak, and QCD interactions \cite{PRL}.}
	\label{fig:feynman}
\end{figure}

The motivation for muon $g-2$ experiments comes from various reasons. First of all, the muon anomaly $a_\mu \equiv \frac{g-2}{2}$ is a relatively simple observable to measure. The $g$ factor relates the magnetic moment of the muon with its spin:
\begin{equation}\label{eq:mu_spin}
\vec{\mu} = g\frac{q}{2m}\vec{S}.
\end{equation}
Such measurements can be obtained through a spin-polarized beam of muons, an external and measurable magnetic field, and a technique to determine the spin precession through time. While such a setup is easy to create with current technology, many challenges arise when trying to achieve an accuracy of less than 1 part-per-million (ppm).\\
Despite the difficulties, we pursue precision measurements of $a_\mu$ because the anomaly probes all of the interactions between the lepton and virtual particles. The whole Standard Model is tested, since all possible particles can contribute either via direct virtual interaction with the muons or with higher-order loop corrections, adding to the value of $a_\mu$. Figure \ref{fig:feynman} depicts the lower-order interactions for the electromagnetic, weak, and strong forces. Table~\ref{table:a_mu_values} lists the contributions to $a_\mu$ by the three SM forces.\\
Moreover, eventual measurements that differ from the SM theoretical prediction would indicate hints of \textit{new physics}. Such results would help setting new constraints on the multitude of currently proposed Beyond Standard Model (BSM) theories.\\
Muons are particularly important in such a measurement because BSM interactions with massive particles contribute with mass suppression terms, $\propto (\frac{m_{lepton}}{M})^2$. The electron $g-2$ has already been measured with around $\mathcal{O}(10^3)$ times more precision than the muon \cite{Hanneke}, but the relative mass ration between the electron and the muon enhances the sensitivity to these terms by a factor of $(\frac{105.66}{0.511})^2\approx 43000$ \cite{Jegerlehner}. Following the same reasoning, $\tau$ leptons would be even better for detecting new heavy particles, but their very short lifetime $(2.9\times10^{-13}$ s), multiple decay branches, and other practical reasons make such an experiment not feasible with current technology.

\begin{table}[!ht]
	\centering
	\caption{Standard model contributions to $a_\mu$ \cite{WP2020}.}
	\label{table:a_mu_values}
	\begin{tabular}{lr}
		\hline
		Term&Value ($\cdot10^{-11}$)\\\hline
		$a_\mu^{QED}$&$116\,584\,718.931\pm0.104$\\
		$a_\mu^{EW}$&$153.6\pm1.0$\\
		$a_\mu^{HVP}$&$6\,845\pm40$\\
		$a_\mu^{HLBL}$&$92\pm18$\\\hline
		Total SM&$116\,591\,810\pm43$\\\hline
	\end{tabular}
\end{table}

\section{The experiment}\label{sec:experiment}
The new Muon $g-2$ Experiment in operation at Fermi National Accelerator Laboratory aims to measure the muon's anomalous magnetic moment with a precision of 0.14 ppm, a factor of four better than the previous BNL E821 Experiment. To achieve this precision, a statistics of $\sim$20 times the amount collected at BNL is needed, and the systematics must be contained within 100 ppb. In order to obtain a large number of observed muons, the BNL storage ring was moved from Brookhaven to Fermilab and installed in the FNAL muon campus accelerator chain, where the beam is cleaner and much more intense.\\
The experimental technique consists of producing a polarized and clean beam of muons, sending it to a storage ring with very uniform magnetic field, and observing the decay positrons. Three measurements are needed to calculate the muon anomaly $a_\mu$: the muon anomalous precession frequency, the magnetic field intensity, and the beam distribution inside the storage region. The master formula is the following:
\begin{equation}
a_\mu = \frac{\omega_a}{\tilde\omega_p}\frac{g_e}{2}\frac{m_\mu}{m_e}\frac{\mu_p}{\mu_e}~,
\label{eq:amu:a_mu_2}
\end{equation}
where $\omega_a$ is the muon anomalous precession frequency, and $\tilde{\omega}_p$ is the Larmor precession frequency of the proton ($\omega_p$) convoluted with the beam distribution, representing the average field intensity experienced by the muons. The remaining factors are known with sufficient precision from other experiments.
\subsection{The muon beam}\label{sec:beam}
\begin{figure}[!ht]
	\centering
	\includegraphics[width=0.8\linewidth,clip]{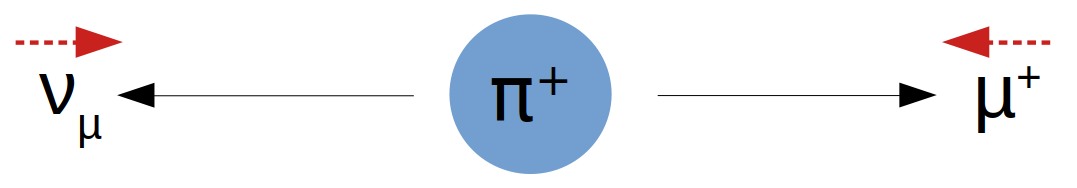}
	\caption{Decay of the positive pion in the rest frame of reference. The spin of the muon points toward the pion as the neutrino must be left-handed because of parity violation of the weak interaction.}
	\label{fig:pion_decay}
\end{figure}

The muons observed in the Muon $g-2$ Experiment originate from decaying pions, which are in turn produced by the collisions of an 8 GeV proton beam on an \textit{Inconel}\textregistered~600 target. After the collisions, secondary particles are focused with an electrostatic lithium lens, and pions having 3.1 GeV momentum are extracted. The beam of pions then circulates inside a Delivery Ring, where remaining protons get removed and all the pions decay into muons. Since pions have zero spin, the muons are emitted isotropically in the rest frame of reference, but their helicity is constrained by the weak decay, as illustrated in Fig.~\ref{fig:pion_decay}. By selecting boosted muons with higher momentum, it is possible to obtain a highly polarized beam in the laboratory frame of reference.
Finally, the 120-ns long muon bunches enter the Muon $g-2$ storage ring at an instantaneous rate of 100 Hz (11.4 Hz average). A superconducting inflector magnet lets the beam pass through the ring yoke into the storage region. Then, after the first quarter of an orbit, a fast pulsed kicker magnet deflects the muon bunch into the final orbit, which has a radius of 7.112 m and a period of 149.2 ns. The $\sim$5000 stored muons per bunch then circulate inside the ring for 700 $\mu$s, while a set of electrostatic quadrupoles provides weak focusing for vertical confinement.

\subsection{The field measurement}\label{sec:field}
The magnet field measurement consists of various components and techniques used to determine the value of $\omega_p$: 378 Nuclear Magnetic Resonance (NMR) fixed probes are placed along the ring under and over the vacuum chamber. The probes keep monitoring the field during the whole data taking. Every three days during data taking, trolley runs are performed in which a mobile cylinder equipped with 17 NMR probes traverses the beam storage region to make 9000 measurements along the ring \cite{Field}. The trolley measurements produce a three-dimensional (3D) map of the magnetic field inside the storage region. The 3D map is then interpolated over the time between two consecutive trolley runs with the fixed probes data.
While the magnetic field is ppm-level uniform along the beam cross section, it is necessary to know the muon distribution inside the storage ring in order to achieve an uncertainty of 70 ppb on $\tilde{\omega}_p$. For this, a set of two tracker stations observe positron tracks and extrapolate the decay vertex in the storage orbit. The beam distribution is extrapolated over the rest of the ring with data from the calorimeters and multiple dedicated simulations \cite{Beam}.

\subsection{Muon precession frequency measurement}\label{sec:omega_a}
A charged particle with mass $m$ placed in an uniform external magnetic field will follow a circular path because of the the Lorentz force, and this motion is called cyclotron motion. If the particle has spin, the spin direction will also rotate (precess) around the direction of the magnetic field.
In the absence of electrical fields, and with the particle velocity and spin perpendicular to the magnetic field, the equations governing this motion are:
\begin{align}
\vec{\omega}_s &= -\frac{ge\vec{B}}{2m} - (1-\gamma)\frac{e\vec{B}}{m\gamma}\label{eq:amu:omegas}\\
\vec{\omega}_c &= -\frac{e\vec{B}}{m\gamma}\label{eq:amu:omegac} ~,
\end{align}
where $\omega_s$ is the spin precession angular frequency and $\omega_c$ is the cyclotron frequency. The second term of Eq.~\ref{eq:amu:omegas} is the Thomas correction to the Lorentz force, which accounts for the rotation of the particle's frame of reference. Subtracting Eq.~\ref{eq:amu:omegac} from Eq.~\ref{eq:amu:omegas}, we obtain the rotation of the particle spin with respect to its momentum. This is the anomalous precession frequency $\omega_a$:
\begin{equation}
\vec\omega_a \equiv \vec\omega_s-\vec\omega_c \simeq -\left(\frac{g-2}{2}\right)\frac{e\vec{B}}{m} \equiv -a_\mu \frac{e\vec{B}}{m} ~.
\label{eq:amu:omega_a}
\end{equation}
However, the particle beam is not always perfectly parallel to the storage plane, but oscillates and breathes vertically and horizontally. The anomalous precession frequency is sensible to such effects:
\begin{align}
\vec{\omega}_a = &-\frac{e}{m}\Bigg[a_\mu\vec{B} - \left(a_\mu - \frac{1}{\gamma^2-1}\right)\frac{\vec{\beta}\times\vec{E}}{c}\nonumber\\
&-a_\mu\left(\frac{\gamma}{\gamma+1}\right)(\vec{\beta}\cdot\vec{B})~\vec{\beta}~\Bigg] ~.
\label{eq:amu:omega_a_magic}
\end{align}
The storage region houses several electrostatic quadrupoles for a weak vertical focusing of the beam. This would affect the muon precession frequency, but, for a muon beam with \textit{magic} momentum $p_\mu$ = 3.094 GeV/c, corresponding to a value of $\gamma = 29.3$, the second term of Eq. \ref{eq:amu:omega_a_magic} cancels out. The remaining effect and the third term are measured separately and applied as \textit{E-field} and \textit{Pitch} corrections to $\omega_a$.\\

\begin{figure}[!ht]
	\centering
	\includegraphics[width=0.8\linewidth,clip]{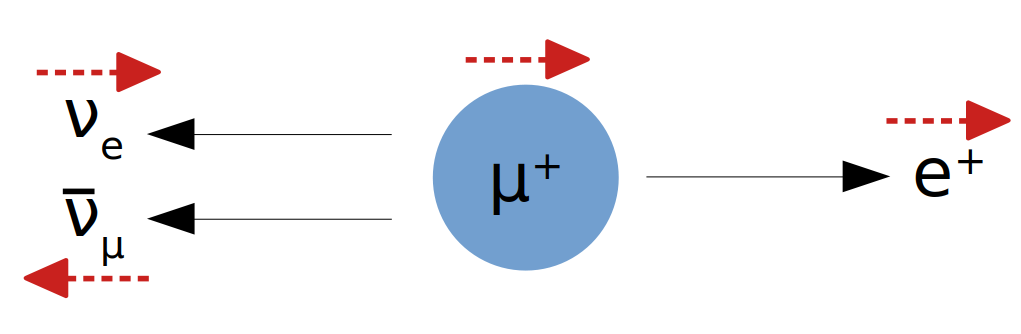}
	\caption{Muon decay configuration that maximizes the positron momentum and decay probability. The right-handed positron is emitted in the direction of the muon spin.}
	\label{fig:muon_decay}
\end{figure}

\begin{figure}[!ht]
	\centering
	\includegraphics[width=0.9\linewidth,clip]{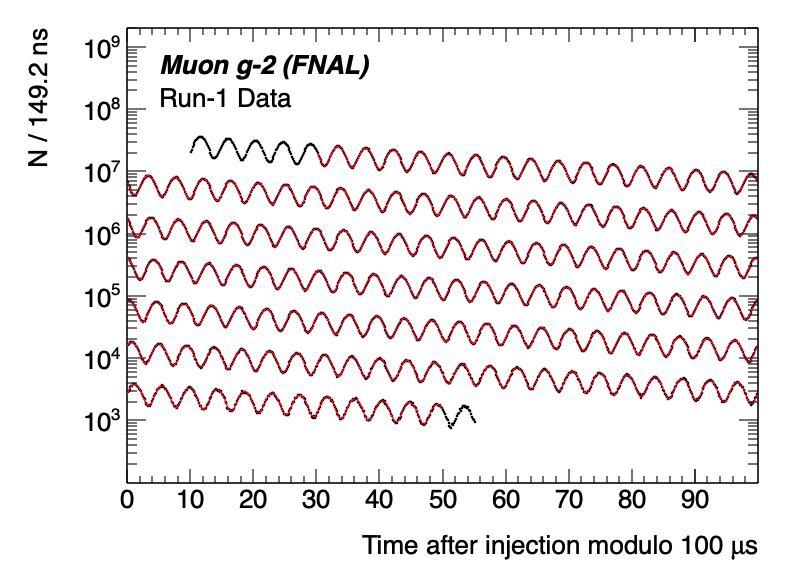}
	\caption{\textit{Wiggle plot} showing the number of $>$1.7 GeV positrons over time. The graph is wrapped every 100 $\mu$s to show the entire measurement period. The oscillation period is $2\pi/\omega_a$.}
	\label{fig:wiggle}
\end{figure}

The measurement principle relies on the parity-violating nature of the weak decay. The positive muons decay into a positron and two neutrinos with nearly $100\%$ probability. In the rest frame of the muon, the highest energy decay positrons come from decays in which the neutrinos are emitted collaterally, as depicted in Fig.~\ref{fig:muon_decay}. In this scenario, half of the initial rest mass of the muon is carried away by the decay positron ($E_{max} \approx 53$ MeV), while the rest is shared by the two neutrinos. Since the neutrino and anti-neutrino are traveling in the same direction, and the weak decay dictates they must have opposite helicities, their spins must be opposite. With the neutrinos' spins canceling, conservation of angular momentum forces the decay positron to carry the spin of the parent muon.
The $V-A$ nature of the weak decay prefers to couple to a right-handed positron, so the high-energy decay positron depicted in Fig.~\ref{fig:muon_decay} tends to be emitted in the direction of the muon spin. Therefore, in the rest frame of the muon, the spin direction of the muon can be monitored by observing the direction at which the high energy decay positrons are emitted.
The Muon $g-2$ Experiment is equipped with 24 electromagnetic calorimeters located around the ring with 15° azimuthal distance between each other. The asymmetry of the decay process, together with the fact that the spin precesses with respect to the momentum, results in an oscillation in the count of positrons over time. The number of detected positrons above a single energy threshold $E_{th}$ is:
\begin{equation}
N(t) = N_0 e^{-t/\tau}[1+A\cos(\omega_a t + \phi)] ~,
\label{eq:5par}
\end{equation}
where the normalization $N_0$, the asymmetry $A$ and the initial phase $\phi$ are all dependent on the energy threshold. $\tau$ represents the lifetime of the muon in the laboratory frame of reference, that is $\gamma \tau_\mu \simeq 64.4$ $\mu$s. Additional beam-related effects, as coherent betatron oscillations and muon losses, appear as multiplicative terms to $N_0$, $A$, and $\phi$. Three different reconstruction techniques and several analysis procedures are implemented to obtain the $\omega_a$ value \cite{Omega}. As an example, in the A-Weighted method the positrons counts are weighted by the asymmetry, which depends on the energy, yielding the maximum possible statistical power for a given threshold $E_{th}$. The typical plot showing the count of positrons over time is called \textit{wiggle plot} and is shown in Fig. \ref{fig:wiggle}.

The positrons are measured by 24 calorimeters, which each have a 9x6 array of PbF$_2$ crystals with size $2.5\times2.5\times14$ cm. The charged particles in the electromagnetic shower generate \v{C}erenkov photons. The choice of a pure \v{C}erenkov material is driven by the almost instantaneous signal produced when a positron strikes, further enhanced by a black Tedlar\textregistered~wrapping that absorbs part of the reflected photons. Every crystal is coupled with a Silicon PhotoMultiplier (SiPM) detector, whose signals are digitized at an 800 MHz sampling rate and chopped into 40-ns islands by on-line GPUs. A precise gain calibration of the SiPMs at the level of $\mathcal{O}(10^{-4})$ is provided by a Laser-based Calibration System \cite{Laser}.

\subsection{Blinding}\label{sec:blinding}
The Muon $g-2$ Experiment has an unique feature, which is the hardware blinding of the digitization clocks. Before the start of each run of data acquisition, the main clock that determines the calorimeters digitization rate is configured with a secret offset in the 25 ppm range. The real frequency is known only by two persons outside of the collaboration, and that number is only revealed to the analyzers at the final stage of the measurement process. The whole muon precession analysis is therefore performed blindly on time-uncalibrated data. When the collaboration unanimously decides that the analysis is mature and the final value is calculated, the secret frequency is revealed.

\section{Status and future of the Muon $g-2$ Experiment}\label{sec:status}

\begin{figure}[!ht]
	\centering
	\includegraphics[width=0.9\linewidth,clip]{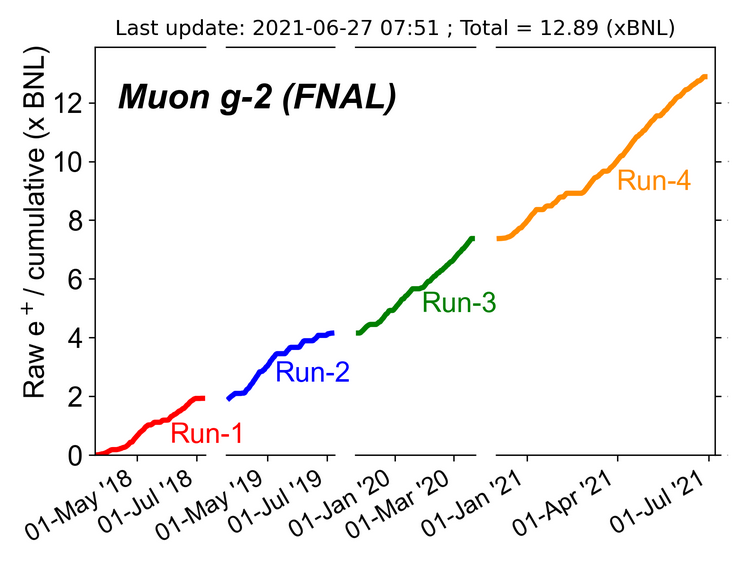}
	\caption{Collected statistics compared to the BNL dataset in the first four runs of the experiment.}
	\label{fig:ctags}
\end{figure}

The Muon $g-2$ Experiment at Fermilab is currently running its fifth year of data taking. In the first four years, an equivalent statistics of 13$\times$BNL has been collected (Fig. \ref{fig:ctags}).
The Run-1 data has been completely analyzed and has been published earlier this year. The result was in agreement with the BNL experiment and the combined value increased the discrepancy with theory to 4.2 $\sigma$ \cite{PRL}.\\
The data from Run-2 through Run-4 data is currently under reconstruction and analysis, with many efforts in improving the systematic errors with respect to Run-1, both on the hardware and the software fronts. For instance, the magnet has been covered with a thermal blanket, and the hall is now temperature-controlled with air conditioners. The kicker strength has been improved~\cite{Kicker}, and the electrostatic quadrupole resistors have been upgraded.
New beam scraping techniques and quadrupole radio-frequency pulses have been put in place to increase the storage efficiency and minimize the betatron oscillations \cite{QuadRF}. On the analysis side, independent efforts from the $\omega_a$ groups try to increase pileup separation and decrease gain related systematics. The beam-related effects to $\omega_a$ are being calculated with improved precision thanks to ever increasing simulation studies. The Run-5 data taking just started in late 2021, and another Run-6 is currently planned. The collaboration is considering the possibility of using negative muons for the Run-6 data taking. Such a run would provide an opportunity to test the robustness of the method and verify CPT and Lorentz conservation.

\section{Acknowledgments}
This work was supported in part by the US DOE, Fermilab, the Istituto Nazionale di Fisica Nucleare (Italy), and the European Union's Horizon 2020 research and innovation programme under the Marie Sk\l odowska-Curie grant agreements No. 690835 (MUSE) and No. 734303 (NEWS).

\end{document}